\documentclass[a4paper,11pt]{article}

\usepackage{fullpage}
\usepackage[T1]{fontenc}
\usepackage{xcolor,cite,hyperref,caption}
\usepackage{amssymb,amsmath,mathtools}
\usepackage{graphicx,ulem}

\newcommand{\be}{\begin{eqnarray}}
\newcommand{\ee}{\end{eqnarray}}

\allowdisplaybreaks

\numberwithin{equation}{section}

\begin{document}

\setcounter{footnote}{0}

\baselineskip 6 mm

\begin{titlepage}
	\thispagestyle{empty}
	\begin{flushright}
		
	\end{flushright}

	\vspace{35pt}
	
	\begin{center}
	    { \LARGE{\bf Scale-separated vacua with extended supersymmetry}} 
		
		\vspace{50pt}
		
		{Niccol\`o Cribiori$^1$,  Fotis Farakos$^2$,  Alexandros Zarafonitis$^2$}  
		
		\vspace{25pt}

        $^1${\it KU Leuven, Institute for Theoretical Physics, \\ 
        Celestijnenlaan 200D, B-3001 Leuven, Belgium}
		
		\vspace{15pt}

        $^2${\it Physics Division, National Technical University of Athens \\
        15780 Zografou Campus, Athens, Greece}
		
		\vspace{15pt}

		\vspace{40pt}
		
		{ABSTRACT} 
	\end{center}

We propose the first examples of scale-separated vacua with extended supersymmetry. They arise as circle compactifications of four-dimensional vacua  of massive type IIA supergravity with scale separation, upon introducing additional fluxes and sources. We provide both the ten-dimensional solutions and the three-dimensional effective descriptions in terms of K\"ahler potential and superpotential. The conformal dimensions of the putative dual two-dimensional field theory appear not to be integers. The superpotential for the additional fluxes of one of our models was guessed by \texttt{ChatGPT} and, to the best of our knowledge, it does not appear in existing literature. Should these vacua be solutions of string theory, they would allow to address the open problem of scale separation from the vantage point of extended supersymmetry.

\vspace{10pt}

\bigskip

\end{titlepage}

\newpage

\tableofcontents

\newpage

\section{Introduction}

An open problem in fundamental physics is to explain why the universe appears four-dimensional from the viewpoint of a macroscopic observer. The problem is particularly interesting in string theory, which is most naturally formulated in ten spacetime dimensions. 
A popular route to bridge this gap is that of string compactifications. They are characterized by a scale, the Kaluza--Klein length, which must be much smaller than the typical length scales of the non-compact universe. Solutions of the equations of motion with this property are said to exhibit scale separation. The vast majority of the existing research in this topic focuses on supersymmetric anti-de Sitter vacua (AdS). This is mainly due to technical reasons: AdS vacua can host supersymmetry which makes them more accessible. 
It turns out, however, that scale-separated vacua of this kind are extremely rare; see \cite{Coudarchet:2023mfs,VanRiet:2023pnx} for reviews and \cite{VanHemelryck:2024bas,VanHemelryck:2025qok,Miao:2025rgf,Tringas:2025bwe,Tringas:2026ncg} for more recent constructions.
Given that they are also under debate, one should remain open to the possibility that none of them may survive deeper scrutiny.

Part of the difficulty in constructing scale-separated vacua lies in the absence of a preferred top-down strategy to realize them. On the other hand, in recent years several bottom-up, holographic and supergravity arguments have been put forward to constrain the existence of such vacua, under suitable assumptions \cite{Apruzzi:2019ecr,DeLuca:2021mcj,Cribiori:2022trc,Montero:2022ghl,Cribiori:2023ihv,Cribiori:2023gcy,Cribiori:2024jwq,Cribiori:2026btb}. As a result, large portions of theory space are believed to be unlikely to support scale separation in supersymmetric AdS vacua. Still, on top of the few four-dimensional models, some corners remain open, notably two-dimensional vacua with minimal supersymmetry and three-dimensional vacua with any amount of preserved supersymmetry. Nevertheless, in the latter case, all known models preserve at most minimal supersymmetry \cite{Farakos:2020phe,Emelin:2022cac,VanHemelryck:2022ynr,Farakos:2023nms,Farakos:2023wps,Bobev:2023dwx,VanHemelryck:2024bas,VanHemelryck:2025qok,Miao:2025rgf,Tringas:2025bwe,Bobev:2025yxp,Tringas:2026ncg}. Indeed, no scale-separated vacuum with extended supersymmetry is known to date, in any number of spacetime dimensions. In this note, we take first steps towards filling this gap.

We present two explicit models of scale-separated three-dimensional AdS vacua with $\mathcal{N}=2$ supersymmetry. They are obtained from the well-studied four-dimensional models of \cite{DeWolfe:2005uu}, which we refer to as DGKT, and \cite{Camara:2005pr}, which we refer to as CFI, by compactifying on a circle and introducing suitable additional flux and source contributions. These models arise as compactifications of massive IIA supergravity on an orientifold of a Calabi--Yau threefold with $O6$-planes. We consider them in their toroidal orbifold limit and include an additional circle that is not affected by the orientifold involution.

To show that these models admit solutions of the equations of motion with extended supersymmetry, while exhibiting scale separation, we proceed as follows. First, we determine a solution of the ten-dimensional Killing spinor equations, together with the ten-dimensional Bianchi identities, which is compatible with at least one Killing spinor. This implies that such solution preserves at least $\mathcal{N}=1$ supersymmetry in three spacetime dimensions, without excluding the presence of additional Killing spinors. Solving the ten-dimensional supersymmetry equations guarantees that the actual equations of motion are satisfied, since any solution of the former is, at least locally, an extremum of the action. We then confirm by direct computation that the solution does exhibit scale separation.
In a second step, we show that the specific solution in fact preserves $\mathcal{N}=2$ supersymmetry. We provide the effective K\"ahler potential and superpotential and we match the $\mathcal{N}=2$ AdS vacuum obtained from the three-dimensional supersymmetric effective action with the one solving the ten-dimensional equations of motion, finding precise agreement. Indeed, the single gravitino of the four-dimensional theory now becomes two gravitini in the three-dimensional theory, therefore unavoidably the description would be within the premises of three-dimensional $\mathcal{N}=2$ supergravity. More precisely, our models feature $\mathcal{N}=(1,1)$ supersymmetry and thus no continuous R-symmetry. To conclude the argument, we show that the same vacuum cannot be obtained within gauged supergravity, thus excluding partial supersymmetry breaking; this also implies that the very same vacuum is not captured by the analysis of \cite{Arboleya:2024vnp,Arboleya:2025jko}.

To the best of our knowledge, our models are the first candidates combining scale separation with extended supersymmetry. Since the vacua are AdS, a natural question would be to understand which properties the putative dual conformal field theories have. The hope is that holography could give a better understanding of these vacua thanks to the presence of non-minimal supersymmetry. As a preliminary investigation, we calculated the conformal dimensions of the operators dual to the bulk (untwisted) scalar fields. For DGKT and CFI in four dimensions, these are found to be integer \cite{Conlon:2021cjk,Apers:2022zjx,Apers:2022tfm}. In our models, instead, all conformal dimensions appear not to be integer.

\section{First model: DGKT on a circle}

We consider massive type IIA supergravity compactified on an orientifold of the toroidal orbifold $T^6/\mathbb{Z}_3^2$, following DGKT \cite{DeWolfe:2005uu}. This construction gives rise to four-dimensional $\mathcal{N}=1$ AdS vacua, which provided the first example exhibiting scale separation and full moduli stabilization. We further compactify on a circle, so that the internal seven-dimensional space becomes
\begin{equation}
\label{eq:7Dintspace1}
X_7= \frac{T^6}{\mathbb{Z}_3\times \mathbb{Z}_3} \times S^1 \, .
\end{equation}
The orientifold involution acts on the $T^6$ coordinates $z_i = x_i + i y_i$, $i=1,2,3$, as
\begin{equation}
\label{eq:sigmaO6}
\sigma_{O6}: \quad z_i \to \bar z_i \, ,
\end{equation}
while leaving the additional $S^1$ coordinate invariant, lying along the $O6$-plane worldvolume. Notice that this differs by a minus sign from the original DGKT involution \cite{DeWolfe:2005uu}. As pointed out in \cite{Junghans:2023yue}, this modification is required in order for the orientifold to be compatible with the orbifold action. Apart from this change, we will mostly follow the notation and conventions of \cite{DeWolfe:2005uu}, and work in string units. In particular, recall that, besides $F_0$, the solution features $H_3$ and $F_4$. The $H_3$ flux is odd under (the pullback of) $\sigma_{O6}$, while $F_4$ is even \cite{Grimm:2004ua},
\begin{align}
\label{eq:sigmaH3F4}
\sigma_{O_6}:\quad \, H_3 \to - H_3, \qquad F_4 \to F_4.
\end{align}

We first determine that this background admits three-dimensional AdS solutions with moduli stabilization (including axions) and scale separation, upon introducing suitable new fluxes and sources. Then, we show that these solutions preserve $\mathcal{N}=2$ supersymmetry on the vacuum.

\subsection{Setup}

The six-dimensional toroidal orbifold is characterized by a holomorphic $(3,0)$-form $\Omega$ and a real K\"ahler form $J$. In the conventions of DGKT, the holomorphic form is expanded on a basis of $\mathbb{Z}_3^2$-invariant harmonic three-forms given by
\begin{equation}
\begin{aligned}
\alpha_0 = - \sqrt{2}\, 3^{1/4} \big( &dx_1 \wedge dx_2 \wedge dy_3 + dy_1 \wedge dx_2 \wedge dx_3 \\
&+ dx_1 \wedge dy_2 \wedge dx_3 - dy_1 \wedge dy_2 \wedge dy_3 \big) \, ,
\end{aligned}
\end{equation}
and
\begin{equation}
\begin{aligned}
\beta_0 = \sqrt{2}\, 3^{1/4} \big( &dx_1 \wedge dx_2 \wedge dx_3 - dy_1 \wedge dx_2 \wedge dy_3 \\
&- dx_1 \wedge dy_2 \wedge dy_3 - dy_1 \wedge dy_2 \wedge dx_3 \big) \, .
\end{aligned}
\end{equation}
Indeed, recall that the orbifold $T^{6}/\mathbb{Z}_3^2$ has $h^{21}=0$. 
Under the involution $\sigma_{O6}$, which acts as $x_i \to x_i$ and $y_i \to - y_i$, the form $\alpha_0$ is odd while $\beta_0$ is even. In our conventions ${\rm Re}\Omega \propto \alpha_0$ and ${\rm Im}\Omega \propto \beta_0$. 
The K\"ahler form is instead expanded on a basis of $\mathbb{Z}_3^2$-invariant harmonic two-forms 
\begin{equation}
\begin{aligned}
\omega_1 &= 2 (\sqrt{3}\,\kappa)^{1/3} \, dx^1 \wedge dy^1 \, ,\\
\omega_2 &= 2 (\sqrt{3}\,\kappa)^{1/3} \, dx^2 \wedge dy^2 \, ,\\
\omega_3 &= 2 (\sqrt{3}\,\kappa)^{1/3} \, dx^3 \wedge dy^3 \, .
\end{aligned}
\end{equation}
Indeed, these are all odd under the orientifold involution, and the orbifold $T^{6}/\mathbb{Z}_3^2$ has $h^{11}=3$. The real constant $\kappa = \int_{T^6/\mathbb{Z}_3^2} \omega_1 \wedge \omega_2 \wedge \omega_3$ is the triple intersection number, whose explicit value will not be needed in what follows.

The six-dimensional coordinate basis used in DGKT is normalized as
\begin{equation}
\int_{T^2_i} dx_i \wedge dy_i = \frac{\sqrt{3}}{2} \, , \qquad i=1,2,3,
\end{equation}
for each of the three two-tori of the orbifold. For our setup with a seven-dimensional internal manifold, in addition to introducing the extra $S^1$ coordinate, it will be convenient to relabel the one-forms as
\begin{equation}
\label{eq:ytoxbasis}
dy^1 = dx^4 \, , \qquad dy^2 = dx^5 \, , \qquad dy^3 = dx^6 \, , \qquad \eta = dx^7 \, ,
\end{equation}
where the last one is dual to the one-cycle of the $S^1$. We then switch to new coordinates $\hat x^i$, now with $i=1,\dots,7$, obtained from the original ones by a rescaling such that each of them has unit period,
\begin{equation}
\int d\hat x^i = 1 \, .
\end{equation}

The notation is meant to mimic the one commonly used in the description of $G_2$-manifolds.  
However, we cannot really freely use a harmonic basis parametrizing a $G_2$-structure, otherwise we would explicitly break the three-dimensional $\mathcal{N}=2$ supersymmetry down to $\mathcal{N}=1$. 
Instead, we will only employ those specific combinations of $G_2$ harmonic forms that preserve the underlying $SU(3)$-structure with which our space is equipped.

Recall that a $G_2$-manifold is associated to an associative three-form $\Phi$.
From the  inclusion $SU(3) \xhookrightarrow{} G_2$, we can still construct explicitly $\Phi$. Let us define the vielbein one-forms $e^i = r_i d\hat x^i$, with no sum over $i= 1, \dots, 7$; the $r_i$ are radii of the seven torus $T^6 \times S^1$.
With these ingredients and following Proposition 1.8 of \cite{Joyce:2002eb}, we can write
\begin{equation}
\label{eq:PhiOmJ}
\begin{aligned}
\Phi & =  2 \sqrt 2 \, {\rm Re}\,\Omega + \frac{1}{\kappa^{1/3}} \eta \wedge J 
\\
& =e^{456}+e^{135}-e^{234}-e^{126}+e^{147}+e^{257}+e^{367}\\
& = \sum_{i=1}^7 s_i\Phi_i \, ,
\end{aligned}
\end{equation}
where the $s_i$ are scalar fields cubic in the $r_i$ and we choose the basis of orientifold-odd three-forms 
\begin{equation}
\label{eq:Phibasis}
\Phi_i = \{ d \hat x^{456},  d \hat x^{135}, - d \hat x^{234}, - d\hat x^{126}, d\hat x^{147},   d\hat x^{257},   d\hat x^{367}  \}.
\end{equation}
In the same way, again following \cite{Joyce:2002eb}, we can identify the coassociative four-form,
\begin{equation}
\begin{aligned}
\Psi &= 2 \sqrt 2 \, \eta \wedge {\rm Im}\,\Omega + \frac{1}{\kappa^{2/3}} J \wedge J 
\\
&=
e^{1237} + e^{2467}-e^{1567}-e^{3457}+e^{2356}+e^{1346}+e^{1245}\\
&=*\Phi\, ,
\end{aligned}
\end{equation}
and a convenient basis of orientifold-even four-forms is 
\begin{equation}
\label{eq:Psibasis}
\Psi_i = \{d\hat x^{1237}, d\hat x^{2467}, -d\hat x^{1567}, - d\hat  x^{3457}, d\hat  x^{2356}, d\hat  x^{1346}, d\hat  x^{1245}\}.
\end{equation}
We set the shorthand notation $A^{ij} = A^i \wedge A^j$, and with respect to \cite{VanHemelryck:2022ynr} we have $e^6|_{\rm here}=-e^3|_{\rm there}$ and $e^3|_{\rm here}=e^6|_{\rm there}$. One can also calculate that $\Phi \wedge * \Phi = 7 e^{1234567} = 7 {\rm vol_7}$, with ${\rm vol_7}$ the seven-dimensional volume form.

We perform the compactification with the ten-dimensional string-frame metric
\begin{equation}
\label{eq:ds10d}
ds_{10}^2 =  ds_3^2 + ds_7^2 \,, 
\end{equation}
where the toroidal internal space
\begin{equation}
\label{eq:ds7d}
ds_7^2 = r_1^2 (d\hat x^2_1 + d\hat x^2_4)  + r_2^2 (d\hat x^2_2 + d\hat x^2_5) + + r_3^2 (d\hat x^2_3 + d\hat x^2_6) + r_7^2 d\hat x_7^2 
\end{equation}
has only four real moduli; namely we have already set $r_4 = r_1$, $r_5 = r_2$ and $r_6 = r_3$.
To further simply the setup, we aim directly for isotropic solutions and thus assume 
\begin{equation}
r_1 = r_2 = r_3 \equiv r \,.
\end{equation}
This leads to two independent moduli,
\begin{equation}
s_1=s_2=s_3=s_4 = r^3 \equiv s \, , \qquad s_5=s_6=s_7 = r^2 r_7 \equiv t \,, 
\end{equation}
and the seven-dimensional volume in string units is ${\rm vol}(X_7) = r^6 r_7 = s^{4/3} t$.

The untwisted cohomology of our seven-dimensional manifold  \eqref{eq:7Dintspace1} has Betti numbers $b_1=1$, $b_2= 3$ and $b_3= 5$. The latter comes from the fact that DGKT has $b_3^{DGKT}=2$, but in our case we get additional three-cycles in the seven-dimensional manifold by combining the  $b_2^{DGKT}=3$ two-cycles of the DGKT orbifold with the additional circle that we inserted, hence $b_3 = b_3^{DGKT}+b_2^{DGKT}=5$. Under the orientifold involution, $b_3^{DGKT,+}=1$ and $b_3^{DGKT,-}=1$ and $b_2^{DGKT,+}=0$ and $b_3^{DGKT,-}=3$. 
Hence, under the same orientifold involution, our three-cycles split into one even and four odd, namely $b_3^{+}=1$ and $b_3^{-}=4$, while the four-cycles split as $b_4^-=1$ and $b_4^{+}=4$, since the volume form ${\rm vol}_7$ entering Hodge duality is odd.
Furthermore, since we defined seven scalars $s_i$ but for us $b_3\neq 7$, comparing \eqref{eq:PhiOmJ} with the usual $G_2$-holonomy fact that $\Phi = \sum_{i=1}^{b_3} s^i \Phi_i$ \cite{Joyce1996G2I}, the holonomy of the seven-dimensional internal space \eqref{eq:7Dintspace1} cannot be $G_2$, indeed we also have a non-trivial one-cycle.

In hindsight, we made a specific choice of harmonic forms basis \eqref{eq:Phibasis} and \eqref{eq:Psibasis} which is already adapted to the $O6$-planes. Besides $F_0$, the setup involves $H_3$ and $F_4$ fluxes, which are respectively odd and even under the orientifold involution. It is therefore natural to formulate the discussion in terms of a three-form $\Phi$ and a four-form $\Psi$. On dimensional grounds, the natural identification is to associate $\Phi$ with $H_3$ and $\Psi$ with $F_4$. The simplest possibility is thus when $\Phi$ has the same orientifold parity as $H_3$, and similarly $\Psi$ has the same parity as $F_4$. 
One could instead choose ${\rm Im}\,\Omega$ in the definition \eqref{eq:PhiOmJ} of $\Phi$, leading to a mixed orientifold parity. While this would be perfectly consistent from the purely geometric viewpoint, it would make it less natural to expand $H_3$ on the basis of the $\Phi_i$ and $F_4$ on the basis of the $\Psi_i$. In this sense, the way we constructed the harmonic form basis is already correlated with both the orientifold configuration and the string theory setup under consideration. Besides, expanding $H_3$ and $F_4$ on the basis of $\Phi_i$ and $\Psi_i$ will also ensure that the structure of the orbifold is preserved, since those forms are $\mathbb{Z}_3^2$-invariant.

\subsection{Circle compactification}

To compactify DGKT we need to switch-on additional flux along the circle, avoiding that it decompactifies leading us back to the four-dimensional DGKT AdS vacuum. 
Given the orientifold parity, the only option we have is to turn on additional $F_4$ flux, while also making sure that it remains compatible with the $\mathbb{Z}_3^2$ orbifold. 
A simple choice is to take
\begin{equation}
\label{eq:F4dgkt+}
F_4 = f \left( \Psi_5 + \Psi_6 + \Psi_7 \right)  + f' (\Psi_1 + \Psi_2 +\Psi_3 +\Psi_4) \,, 
\end{equation}
where the first piece, multiplied by $f$, is the old DGKT flux, while the second piece, multiplied by $f'$, is a new contribution proportional to $\alpha_0 \wedge dy$ in the SU$(3)$-structure language. One can check that this ansatz is such that $F_4$ is invariant under the $\mathbb{Z}_3^2$ orbifold, while it has the proper parity \eqref{eq:sigmaH3F4} under the orientifold involution. 
Notice that the monomials in \eqref{eq:F4dgkt+} multiplied by $f$ can in principle scale independently, as DGKT has $b_2^{DGKT}=b_4^{DGKT}=3$ independent components of $F_4$ flux. Instead, in order not to break the SU$(3)$-structure, we are forced to treat the remaining terms $\Psi_1+\Psi_2+\Psi_3+\Psi_4$ as a single entity: this is the additional four-cycle with respect to DGKT, needed to get $3+1=4=b_4^+$.

With this choice of $F_4$ flux, the tadpole $\int dF_6 =\int F_4 \wedge H_3 + J_{D2}$ would require the presence of $D2$-branes in order to achieve $\int dF_6 =0$. In turn, this would restrict the $f'$ flux. We can avoid this issue by additionally modifying the $H_3$ flux from the standard DGKT choice in order to engineer
\begin{equation}
F_4 \wedge H_3 = 0 \,, 
\end{equation}
such that the tadpole is canceled with $J_{D2}=0$.
Hence, our ansatz for the $H_3$ flux is 
\begin{equation}
\label{eq:H3dgkt+}
H_3 = h ( \Phi_1 + \Phi_2 +\Phi_3 +\Phi_4  ) + h' \left( \Phi_5 + \Phi_6 + \Phi_7 \right) \,, 
\end{equation}
where the first piece, multiplied by $h$, is the old DGKT contribution, while the second piece, multiplied by $h'$ is proportional to $w_i \wedge dy$ in the SU$(3)$-structure language. One can check that also this ansatz is invariant under the $\mathbb{Z}_3^2$ orbifold, while it has the proper parity \eqref{eq:sigmaH3F4} under the orientifold involution. The three monomials in \eqref{eq:H3dgkt+} proportional to $h'$, can in principle scale independently. Instead,  in order not to break the SU$(3)$-structure the terms $\Phi_1+\Phi_2+\Phi_3+\Phi_4$ must be considered as a single entity: this is the original 3-form inherited from DGKT (where $h^{21}=0$ and $H_3 \sim h \beta_0$) and needed to get $3+1 = 4 = b_3^{-}$.

Given that $F_2=0$, the $dF_4$ Bianchi identity is automatically satisfied without the need for $D4/O4$-sources and especially without constraining $F_4$. Since the only other tadpole in which $F_4$ appears is the one stemming from the $dF_6$ Bianchi identity, that we have already discussed and that is solved by $F_4 \wedge H_3=0$,  we basically do not have restrictions on the overall rescaling of $F_4$. Thus, $F_4$ is an unbounded flux and we will parametrize it as
\begin{equation}
\label{eq:ff'Ntoinfty}
|f'| \sim |f| \sim N \,,
\end{equation}
where the assumption $|f'| \sim |f|$ will be motivated in a while. 
The goal will be to find solutions for which the limit $N \to \infty$ corresponds to large volume, weak coupling and scale separation.

The fact that we have extra $H_3$ flux means we need to cancel additional tadpoles with respect to DGKT.
We do not introduce $NS$ sources in such a way that we keep $\int d H_3=0$. Still, $H_3$ enters the various RR tadpoles.  In particular, it contributes to the $dF_2$ Bianchi identity with new terms with respect to DGKT. 
Crucially we cannot amend to this by invoking more O-planes, because they are dictated by the orbifold and moreover supersymmetry will be reduced further. 
We thus insert $D6$-branes in such a way that they are mutually supersymmetric with the pre-existing orientifold planes: in our setup (toroidal orbifold with no worldvolume fluxes) the empirical rule is that the number of mixed Neumann-Dirichlet boundary condition between any two directions should be 4 \cite{Bergshoeff:1997kr}. 
To cancel the new $H_3$ contribution to the $dF_2$ tadpole, the $D6$-branes should be transverse to the $h'$ components in \eqref{eq:H3dgkt+}, hence to the $S^1$ direction as well. 
As byproduct, the pull-back of the $H_3$ flux on the D6-branes vanishes, so that we do not have to worry about Freed-Witten anomalies. 
The positions of the $D6$-branes and the pre-existing $O6$-planes is presented in table \ref{table:D6O6conf}.

\begin{table}[ht]
\centering
\begin{tabular}{c c c c c c c c c c c}
\hline
  &  \multicolumn{3}{c}{AdS$_3$} & $S^1$ & 1 & 2 & 3 & 4 & 5 & 6 \\
\hline
O6 & $\times$ & $\times$ & $\times$ & $\times$ & $\times$ & $\times$ &$\times$ & $-$ & $-$ & $-$ \\
O6 & $\times$ & $\times$ & $\times$ & $\times$ & $\times$ & $-$ & $-$ & $-$ & $\times$ & $\times$ \\
O6 & $\times$ & $\times$ & $\times$ & $\times$ & $-$ & $\times$ & $-$ & $\times$ & $-$ & $\times$ \\
O6 & $\times$ & $\times$ & $\times$ & $\times$ & $-$ & $-$ & $\times$ & $\times$ & $\times$ & $-$ \\
D6 & $\times$ & $\times$ & $\times$ & $=$ & $\times$ & $\times$ & $=$ & $\times$ & $\times$ & $=$ \\
D6 & $\times$ & $\times$ & $\times$ & $=$ & $\times$ & $=$ & $\times$ & $\times$ & $=$ & $\times$ \\
D6 & $\times$ & $\times$ & $\times$ & $=$ & $=$ & $\times$ & $\times$ & $=$ & $\times$ & $\times$ \\
\hline
\end{tabular}
\caption{$D6/O6$ configuration which preserves ${\cal N}=2$ supersymmetry in three dimensions. The ``$\times$'' stands for directions that are threaded by the extended objects, while the ``$-$'' or ``$=$'' for directions on which the extended object is localized (with smeared backreaction). Orientifolds are located at the fixed loci of the $\sigma_{O6}$ involution, which means that the ``$-$'' is either $0$ or $1/2$. D-branes, instead, are not fixed points and their position is dictated by local supersymmetry effects. This means that the ``$=$'' corresponds to some position of the $D6$-branes which could be spread out or form a stack, depending on the extremization of the position moduli. Therefore, $D6$-branes do not have to intersect among themselves and neither with the $O6$-planes. The orientifold planes in the DGKT construction are known not to intersect once the orbifold singularities are blown-up \cite{Junghans:2023yue}.} 
\label{table:D6O6conf}
\end{table}

The $dF_2$ tadpole becomes thus
\begin{equation}
\label{eq:dF2tad}
0=\int d F_2 = \int H_3 \wedge F_0 + J , \qquad \text{with} \qquad J = J_{O6} + J_{D6},
\end{equation} 
where the smeared currents of the $D6$-branes and $O6$-planes are
\begin{equation}
\begin{aligned}
J_{D6} &= Q_{D6} \left( \Phi_5 + \Phi_6 + \Phi_7 \right)   ,  \\
J_{O6} &= Q_{O6} \left( \Phi_1 + \Phi_2 +\Phi_3 +\Phi_4 \right) \, ,
\end{aligned}
\end{equation}
with $Q_{D6}>0$ while $Q_{O6}<0$. 
This condition implies that
\begin{equation}
\label{eq:hh'sign}
h h' < 0 \,, 
\end{equation}
since the two different contributions to the $H_3$ flux should cancel objects with opposite charge. 
Furthermore, the requirement $F_4 \wedge H_3 = 0$ translates to
\begin{equation}
\label{eq:fh}
4 f' h + 3 f h' = 0 \,, \qquad i.e. \qquad h' = - \frac{4f'}{3f} h,
\end{equation}
implying that we have to ask 
\begin{equation}
\label{eq:ff'sign}
f f' > 0 \,. 
\end{equation}
Denoting $F_0=m$, to cancel the $dF_2$ tadpole we also need
\begin{equation}
\label{eq:hmsign}
hm >0 \qquad \text{and} \qquad h'm <0,
\end{equation}
stemming from the fact that the first term has to cancel against the negative O-plane charge, while the second against the positive D-brane charge.

From the condition \eqref{eq:fh} we see that,  when the scaling \eqref{eq:ff'Ntoinfty} is implemented and the limit $N\to \infty$ is taken, $h$ and $h'$ can assume any integer (in fact rational) value, including those required by the consistency conditions of the setup. Furthermore, we also see that assuming $|f| \sim |f'|$ avoids the need to let the source charges scale with $N$. This would indeed be unphysical for $Q_{O6}$, which is a fixed number in the compactification. It might still be possible to relax $f \sim f'$, in such a way that $h$ or $h'$ pick up a dependence on $N$, and yet satisfy the $dF_2$ tadpole by letting $Q_{D6}$ scale with $N$ (and perhaps also $m$, as long as it grows with $N$), but we do not explore this possibility here.

\subsection{Ten-dimensional solution}
\label{sec:10dsolDGKT}

Having discussed how the Bianchi identities and tadpoles are satisfied in our configuration, it remains to actually solve the equations of motion of ten-dimensional supergravity. Looking for a supersymmetric solution allows to solve first-order equations instead of the full, second-order Einstein equations. These first order equations are found by setting to zero the supersymmetry variations of the fermions. A standard trick consists in recasting them into equivalent equations for differential forms instead of spinors, by exploiting the Clifford map. The form of these first-order equations suitable for our context, which are sometimes called bispinor equations, can be found $e.g.$ in \cite{VanHemelryck:2022ynr}; we report them below for convenience.

For the NS sector, we have
\begin{equation}
\label{eq:eomH3}
d \left(e^{2A-\phi} \Phi\right)=0\,, \qquad \Phi \wedge H_3=0\, ,
\end{equation}
where $A$ is a warp factor and $\phi$ the dilaton.
For the RR sector, we have
\begin{align}
\label{eq:eomF6}
e^{3A} * F_6 &= -d\left(e^{3A-\phi}\right)\,,\\
\label{eq:eomF4}
e^{3A} * F_4 &= - e^{3A-\phi}H_3 - 2\mu e^{2A-\phi}\Phi\,,\\
\label{eq:eomF2}
e^{3A} * F_2 &= d\left(e^{3A-\phi}*\Phi\right)\,,\\
\label{eq:eomF0}
e^{3A} * F_0 &= e^{3A-\phi}H_3 \wedge * \Phi +2 \mu e^{2A-\phi}{\rm vol}_7\,,
\end{align}
where $\mu^2 = L_{AdS}^{-2}$. Furthermore, we have the additional constraint
\begin{equation}
\label{eq:bispinorpairing}
\Phi \wedge F_4 - F_0 {\rm vol}_7 = -4 \mu e^{-A-\phi}{\rm vol}_7.
\end{equation}

Our setup is particularly simple and thus these equations can be reduced considerably. We look for solutions with constant scalars $e^A = 1$ and $e^\phi = g_s$, and furthermore we have $F_6=0=F_2$. These facts imply that the equations \eqref{eq:eomH3}, \eqref{eq:eomF6} and \eqref{eq:eomF2} are already verified. Indeed, the condition $\Phi \wedge H_3=0$ is automatically implied by our ansatz \eqref{eq:H3dgkt+}. The remaining equations reduce then to the following set of algebraic conditions
\begin{align}
\label{eq:eom1}
&g_s f' s^{2/3} = -ht -2\mu  s t\,,\\
&g_s f t = - h' s^{4/3} -2 \mu s^{4/3}t\,,\\
&g_s m = 4\frac{h}{s} + 3 \frac{h'}{t}+2\mu\,,\\
&4 g_s \frac{f'}{s^{1/3}t} + 3 g_s \frac{f}{s^{4/3}} - g_s m+4\mu=0\,,\\
\label{eq:eom5}
&g_s m = -4 \mu\,,\\
\label{eq:eom6}
&4 \frac{h}{s}+3 \frac{h'}{t} = -6\mu\,,
\end{align}
with the last two being linear combinations of the others. To obtain these relations it is convenient to use $*\Phi_i = \frac{{\rm Vol}(X_7)}{(s_i)^2}\Psi_i $ and $*\Psi_i = \frac{(s_i)^2}{{\rm Vol}(X_7)}\Phi_i$, as well as  $\Phi_i \wedge \Psi_j = \frac{{\rm vol}_7}{{\rm Vol}(X_7)} \delta_{ij}$ and $\Phi_i \wedge * \Phi_j = \frac{{\rm vol}_7}{(s_i)^2}\delta_{ij}$.

As discussed in the previous section, we let $|f|\sim |f'| \sim N$ while keeping fixed $h$, $h'$, together with $m$. From \eqref{eq:eom5} we see that $g_s \sim |\mu| \sim N^{\gamma_1}$, for some $\gamma_1<0$, allows for weak coupling. We then assume that $s \sim t \sim N^{\gamma_2}$, for some $\gamma_2>0$ in order to achieve large volume. From \eqref{eq:eom6}, we find $\gamma_1 = -\gamma_2$ and then \eqref{eq:eom1} fixes $\gamma_1=-\frac34$, $\gamma_2 = \frac34$; one can check that the remaining equations are compatible with this choice. We have thus found a scaling symmetry of the equations of motion
\begin{equation}
|f|\sim |f'| \sim N, \qquad s \sim t \sim N^{\frac34}, \qquad g_s\sim |\mu| \sim 1/L_{AdS} \sim N^{-\frac34}\, ,
\end{equation}
which for $N\to \infty$ leads to the desired properties, namely large volume, weak coupling and parametric scale separation,
\begin{equation}
\label{eq:scalesep}
{\rm Vol}(X_7) \sim N^{\frac74}, \qquad g_s \sim N^{-\frac34}, \qquad \frac{L_{KK}}{L_{AdS}} \sim N^{-\frac12} \, ,
\end{equation}
where we set the Kaluza-Klein length $L_{KK} \simeq {\rm Vol}(X_7)^{\frac17}$. If a solution of the above equations of motion is found, it will have these properties. It will also satisfy the domain wall bound recently derived in \cite{Cribiori:2026btb}, as one can check. Furthermore, notice that our isotropic scaling ansatz is such that $r_i \sim N^{1/4}$, for $i= 1, \dots, 7$, and thus scale separation is really realized with respect to each of the radii separately, namely $r_i/L_{AdS} \sim N^{-1/2}$ for any $i=1,\dots,7$.

We now find one such solution. It is convenient to introduce the following quantities that do not scale with $N$,
\begin{equation}
A = \frac{2 \mu s}{h}  \, , \qquad B = \frac{2 \mu t}{h'} \, , \qquad C = \frac{g_s f'}{2 \mu s^{\frac13}t}  \, , \qquad D =  \frac{g_s f}{2 \mu s^{\frac43}} \, . 
\end{equation}
Inserting these into \eqref{eq:eom1}-\eqref{eq:eom6} and \eqref{eq:fh}, we find
\begin{align}
6A + 8B + 6 AB&=0\,,\\
AC + A + 1 &=0\,,\\
4C + 3D + 4 &=0\,,\\
\frac{A}{B} + \frac43 \frac{C}{D} &=0\,,
\end{align}
whose solution is
\begin{equation}
A = -\frac76\,, \qquad B = 7\,, \qquad C = -\frac17\,, \qquad D = -\frac87\,.
\end{equation}
This fixes all moduli, the string coupling and the AdS radius to the values
\begin{align}
\label{eq:sol10ds}
s &= \frac{7^{\frac34}}{2\sqrt 2}\left(\frac fm\right)^{\frac 34}\, ,\\ 
t &= 2 \sqrt 2\,\, 7^{\frac34} \frac{f'}{m}\left(\frac{m}{f}\right)^\frac14\,,\\ 
\mu &=1/L_{AdS}= -\frac{7^{\frac14}}{3\sqrt 2} h \left(\frac mf\right)^\frac34\, , \\
\label{eq:sol10dgs}
g_s &= \frac{2\sqrt 2 }{3}7^{\frac14} \frac hm \left(\frac mf\right)^{\frac34}.
\end{align}
We have thus found a solution with full moduli stabilization at large volume, weak string coupling and with parametric scale separation as $N \to \infty$.  For the geometric moduli to be real, we have to require that $f$ and $m$ have the same sign. Then, for $g_s$ to be positive, $h$ and $m$ must have the same sign as well. Taking into account also our previous constraints on the various signs, we see that there are two branches of solutions, depending on the choice of, say, $m$. For $m>0$, we need to have $h>0$ and $h'<0$, and in turn $f>0$ and $f'>0$. For $m<0$, we need to have $h<0$ and $h'>0$, and in turn $f<0$ and $f'<0$. The opposite choice for the signs of $f$ and $f'$ would lead to the ``skew-whiffing'' non-supersymmetric solutions that are typically found by flipping signs from supersymmetric ones, see $e.g.$ \cite{DeWolfe:2005uu,Farakos:2020phe}. Indeed, the three-dimensional scalar potential computed in the next section will depend only on $|F_4|^2$ and thus an extremum is found regardless of the sign of $f$, $f'$.

\subsubsection*{Twisted sector}
Let us briefly discuss the twisted sector. With respect to \cite{DeWolfe:2005uu}, our setup features an additional circle compactification which however does not introduce new orbifold singularities. Hence, we have the same nine twisted moduli as DGKT.

Given that we lack an explicit metric for the smooth Calabi-Yau manifold after the resolution, we can only provide a scaling analysis of moduli stabilization in the twisted sector by means of a local approximation. Each of the nine orbifold singularities is locally $\mathbb{C}^3/\mathbb{Z}_3$. Its resolution by a $\mathbb{CP}^2$ is described by the metric \cite{DeWolfe:2005uu,Junghans:2023yue}
\begin{equation}
ds^2 = \rho^2 ds^2_{\mathbb{CP}^2} + F(\rho)^{-1} d\rho^2 + \frac{\rho^2}{9}F(\rho) (d\theta - \mathcal{A})^2 + r_7^2 \eta^2,
\end{equation}
where $F(\rho) = 1-a^6/\rho^6$ with $a$ the modulus of $\mathbb{CP}^2$ and $\rho\geq a$ the radial coordinate measuring departure from the blown-up singularity. $\mathcal{A}$ is a one-form whose explicit expression can be found in \cite{DeWolfe:2005uu, Junghans:2023yue}, while $\eta$ is the one-form associated to the additional circle with radius $r_7$. The volume form is proportional to $\sqrt{g_7}\sim r_7 \rho^5$, and thus the blown-up of the singularity removes a region of radius $a$ and volume proportional to $\int_0^a \rho^5 d \rho \sim a^6$ times $r_7$. This estimate confirms that our additional circle acts only as spectator and thus the local analysis is the same as in DGKT. We denote with ${\rm Vol}_0$ the volume of the six-dimensional singular manifold and with ${\rm Vol} = {\rm Vol}_0 - \beta\, a^6$ the corrected  six-dimensional volume, for some coefficient $\beta$ whose value is not important at this level of precision; the full volume is thus $r_7 {\rm Vol}$.

To stabilize the modulus $a$ we thread $\mathbb{CP}^2$ with $\tilde f$ units of $F_4$ flux. The relevant couplings in the three-dimensional scalar potential are thus $\int F_0 \wedge * F_0 \sim m^2 r_7\int_a^\infty \rho^5 d\rho \sim m^2 r_7 {\rm Vol}=m^2 r_7 ({\rm Vol}_0 - \beta\,\,a^6) $ and $\int F_4 \wedge * F_4 \sim \tilde f^2 r_7\int_a^\infty \rho^5 \rho^{-8} d\rho \sim\tilde f^2 r_7 /a^2$. After going to Einstein frame, they lead to the scalar potential
\begin{equation}
V(a)/r_7 \sim \frac{\tilde f^2}{a^2} + m^2 \beta\,\,a^6  + \dots\, 
\end{equation}
where $\dots$ contain terms not relevant for us now. Minimization with respect to $a$ is achieved for 
\begin{equation}
a \sim \left(\frac{\tilde f}{m}\right)^\frac14\, ,
\end{equation}
precisely as in DGKT. Hence, it is sufficient to require $|\tilde f| \gg |m|$, without the need to scale $\tilde f$ with $N$, in order to achieve stabilization of the blown-up moduli at some fixed value that is also large enough for the supergravity approximation to be valid.

\subsection{Three-dimensional effective theory}

The solution found in the previous section is manifestly $\mathcal{N}=1$ supersymmetric, since it solves the bispinor equations for a $G_2$-structure. We now show that it is actually $\mathcal{N}=2$ by working out the three-dimensional effective theory; this step is particularly well-motivated due to the presence of scale separation. 
We construct the theory starting from the known four-dimensional effective description of DGKT and then performing a further circle compactification. The non-trivial bit will be to understand which new terms arise in the three-dimensional superpotential due to the new flux contributions in \eqref{eq:F4dgkt+} and \eqref{eq:H3dgkt+}.

The four-dimensional $\mathcal{N}=1$ supergravity describing massive type IIA string theory on the orientifold of the toroidal orbifold $X_6 = T^6/\mathbb{Z}_3^2$ contains four chiral multiplets, $S$ and $T_i$, with $i=1,2,3$ \cite{DeWolfe:2005uu}; there are also nine additional chiral multiplets parametrizing the twisted sector, but we do not consider them in this section. One of the four chiral multiplets, $S$, arises from the complex structure sector of the parent four-dimensional $\mathcal{N}=2$ theory with $h^{21}=0$, and contains the (four-dimensional) dilaton, while the triplet $T_i$ describes the geometric moduli of the K\"ahler sector, with $h^{11}=3$. 

We denote ${\rm Re}T_i = r_ir_{i+3}$, $i=1,2,3$,  the volume of the two-cycles in string units and ${\rm Re} S = e^{-\phi}\sqrt{{\rm Vol}(X_6)}$, with ${\rm Vol}(X_6) =\prod_{i=1}^3 r_i r_{i+3}$. 
We also set $t_i = r_i r_{i+3} r_7=s_{i+4}$, $i=1,2,3$, such that ${\rm Re} T_i = t_i/r_7$. Nevertheless, the solution will feature $r_i = r_{i+3}$ since $h^{21} = 0$. 
The imaginary parts needed to complete these multiplets are provided by axions arising from the reduction of $B_2$ on the $b_2^{DGKT,-}=3$ two-cycles and of $C_3$ on the $b_3^{DGKT,+} = 1$ three-cycle. 
We can map to the conventions of \cite{DeWolfe:2005uu} upon multiplying our multiplets by $i$ and using also $ \kappa_{10}={4\pi}/{( 2\pi \sqrt{a'})^8}=1$. 
Eventually, the K\"ahler potential and the superpotential of the four-dimensional supergravity read (up to a real constant) \cite{DeWolfe:2005uu} 
\begin{align}
K_{DGKT} &= -\sum_{i=1}^3 \log (T_i + \bar T_i) - 4 \log (S +\bar S)\, ,\\
\label{eq:WDGKT}
W_{DGKT} &= 2 i h \, S - \frac{i}{2} f \sum_{i=1}^3 T_i + 2 i m \, T_1 T_2 T_3  \,  . 
\end{align}

On this theory, we perform a further circle reduction. Since the circle in the microscopic model is even under the orientifold, the reduction preserves the full amount of supersymmetry, which is four supercharges. The minimal spinor in three dimensions carries two supercharges and thus our three-dimensional theory is $\mathcal{N}=2$ supergravity.  The string frame metric we consider for the whole reduction is 
\begin{equation}
\begin{aligned}
ds^{2}_{10} &= \frac{e^{2\phi}}{{\rm Vol}(X_6)} ds^2_4 + ds_6^2\\
&= \frac{1}{({\rm Re}\, S)^2}\left(\frac{1}{{\rm Re}\,R} ds_3^2+ {\rm Re}\, R \,\,dx_7^2\right) + ds_6^2\, ,
\end{aligned}
\end{equation}
where $R$ is a new chiral superfield with ${\rm Re} \, R = (r_7 \,\, {\rm Re}\, S)^2$ parametrizing the size of the circle and with ${\rm Im }\, R$ the axion obtained by dualizing the graviphoton. Both $ds^2_4$ and $ds^2_3$ are now directly in Einstein frame.
From the metric, we can infer that the three-dimensional K\"ahler potential has to be
\begin{equation}
\label{eq:K3D1}
K = -\log(R+\bar R)- \sum_{i=1}^3 \log (T_i + \bar T_i) - 4 \log (S +\bar S)\, .
\end{equation}
However, using the same superpotential as in \eqref{eq:WDGKT} would not stabilize ${\rm Re }\, R$. We have indeed to supplement \eqref{eq:WDGKT} with additional terms to take into account the new $F_4$- and $H_3$-flux contributions with respect to DGKT. We propose an educated guess for $W$ and then check that the scalar potential it produces matches precisely with the one obtained via direct dimensional reduction. 
Our proposal for $W$ is
\begin{equation}
\begin{aligned}
\label{eq:W3D1}
W &= 2 i h \, S - \frac{i}{2} f \sum_{i=1}^3 T_i + 2 i m \, T_1 T_2 T_3 
\\
& + i \frac{h'}{2} S^2 \left[ \sqrt{\frac{T_2 T_3}{R T_1}} + \sqrt{\frac{T_1 T_2}{R T_3}} + \sqrt{\frac{T_3 T_1}{R T_2}} \right] 
- 2 i f' S \sqrt{\frac{T_1 T_2 T_3}{R}}  \, ,
\end{aligned}
\end{equation}
where the first line is $W_{DGKT}$ while the second contains new terms that consistently vanish in the limit ${\rm Re \, }R \to \infty$.

The three-dimensional scalar potential is found via the standard supergravity formula \cite{deWit:2003ja}
\begin{equation}
V = e^K\left(g^{I \bar J} D_I W \, \bar{D}_{\bar J} \overline W  - 4 W \overline W\right),
\end{equation}
with $g_{I \bar J} = \partial_I \partial_{\bar J}K$ the K\"ahler metric on the scalar manifold. 
Setting the axions to zero, and in the ten-dimensional variables $t_i = r_i r_{i+3} r_7$, $s=r_1r_2r_3$ and $g_s=e^\phi$, the scalar potential reads
\begin{equation}
\begin{aligned}
\label{eq:V3dDGKT}
\left(256 \, s^{8/3} (t_1 t_2 t_3)^{2/3} /g_s^{6}\right) V &= 16 m^2\\
&+\frac{4 h^2}{g_s^2 s^2} + \frac{ (h')^2}{g_s^2}\left(\frac{1}{t_1^2}+ \frac{1}{t_2^2}+\frac{1}{t_3^2}\right)\\
&+\frac{f^2}{s^{8/3} (t_1 t_2 t_3)^{2/3}}\left(t_1^2 + t_2^2 + t_3^2\right) + \frac{4 (f')^2}{s^{2/3}(t_1 t_2 t_3)^{2/3}}\\
&-4 \frac{8hm}{g_s \, s}-\frac{8 h' m}{g_s}\left(\frac{1}{t_1}+\frac{1}{t_2}+\frac{1}{t_3}\right)\\
&+\frac{2}{g_s s^{4/3}(t_1 t_2 t_3)^{1/3}}\left(3 fh' + 4f'h\right).
\end{aligned}
\end{equation}
We now compare it to the direct dimensional reduction, using
\begin{align}
F_0 \wedge * F_0 &= F_0^2 \,{\rm vol}_7\, ,\\
e^{-2\phi}H_3 \wedge * H_3 &=e^{-2\phi} {\rm vol}_7 \left[4\frac{h^2}{s^2}+ (h')^2\left(\frac{1}{t_1^2}+\frac{1}{t_2^2}+\frac{1}{t_3^2}\right)\right]\, ,\\
F_4 \wedge * F_4 &= \frac{{\rm vol}_7}{s^{8/3} (t_1 t_2 t_3)^{2/3}}\left[f^2(t_1^2+t_2^2+t_3^2) + 4 (f')^2 s^2\right],
\end{align}
and the source actions in the smeared approximation
\begin{align}
S_{DBI}^{O6} &\simeq \int \frac{e^{-\phi}}{s} {\rm vol}_7\, ,\\
S_{DBI}^{D6,i} &\simeq \int \frac{e^{-\phi}}{t_i}{\rm vol}_7\,.
\end{align}
We see that the first line of \eqref{eq:V3dDGKT} arises from $F_0 \wedge * F_0 $, the second line from $H_3 \wedge * H_3$, and the third line from $F_4 \wedge * F_4$. 
The fourth line is to be interpreted as contribution from smeared $O6$-planes ($hm$) and $D6$-branes ($h'm$) sources canceling the $dF_2$ tadpole, as required by consistency of the ten-dimensional model. In particular, for this interpretation to be correct, we need to ask $hm>0$ and $h'm<0$, in accordance with \ref{eq:hmsign}. 
The last line is in principle an additional contribution, which looks like a term stemming from smeared $D2/O2$-sources. Since we do not have such sources in the microscopic, this would spoil our matching, unless it vanishes. Indeed, one can see that the condition for it to vanish is precisely the tadpole constraint \eqref{eq:fh}. Upon enforcing it, our effective three-dimensional scalar potential reproduces precisely the ten-dimensional model. We believe that this is non-trivial evidence for the fact that the superpotential \eqref{eq:W3D1} correctly describes the reduction of the ten-dimensional model. 

Next, we look for supersymmetric AdS vacua of the three-dimensional theory. These are found by solving the F-term conditions 
\begin{equation}
D_I W =0 \, .
\end{equation}
A solution is given by ($T_i \equiv T$ and $t_i \equiv t$ for $i=1,2,3$)
\begin{equation}
S = \frac{3 \sqrt 7}{16} \frac{f}{h}\left(\frac{f}{m}\right)^{\frac12}, \qquad T =\frac{\sqrt 7}{4} \left(\frac fm\right)^{\frac12 }, \qquad R =\frac{567 \sqrt 7}{256} \frac{(fh')^2}{h^4} \left(\frac fm\right)^{3/2}\, ,
\end{equation}
or in ten-dimensional language
\begin{equation}
s = \frac{7^{3/4}}{8} \left(\frac fm\right)^{3/4}, \qquad t =  7^{3/4}\frac{f'}{m}\left(\frac mf\right)^{\frac14}, \qquad g_s = \frac{2}{3}7^{1/4}\frac{h}{m}\left(\frac mf\right)^{\frac34}.
\end{equation}
The supersymmetric vacuum energy is
\begin{equation}
V = -4 e^K W \overline{W} = - \frac{2^{12}}{3^8 \,\, 7^2} \frac{h^8 m^4}{f^8 (h')^2}
\end{equation}
with $W = \frac{7 \sqrt 7}{64} if\left( \frac{f}{m} \right)^{1/2}$.
We have thus found an AdS vacuum with $\mathcal{N}=2$ supersymmetry. It arises from the same scalar potential of the ten-dimensional model and the solution matches with \ref{eq:sol10ds}-\eqref{eq:sol10dgs} up to a trivial rescaling of the fields, $s\to 2\sqrt 2 \,s$, $t \to 2 \sqrt 2 \, t$, $g_s \to \sqrt 2\, g_s$. We believe this to provide strong evidence for the existence of scale-separated vacua with extended supersymmetry. 
We have also evaluated the full Hessian matrix on this specific solution and found that it has non-vanishing determinant, implying full moduli stabilization. Recall, however, that in this three-dimensional theory we are discussing only the closed string moduli and we are restricting to the untwisted sector.

Finally, we calculate the conformal dimensions of the operators of the putative dual two-dimensional field theory. Contrary to what happens in DGKT \cite{Conlon:2021cjk,Apers:2022tfm}, we find that they are not integers. 
In particular, we have
\begin{equation}
1 + \sqrt{1+m^2 L_{AdS}^2 } \simeq \left\{ 3.57 , 3.57, 4.57, 4.57, 1.91 , 3.18 , 8.94, 2.91 , 4.18 , 7.94 \right\} \,,
\end{equation}
with the conformal dimension smaller that 2 due to a tachyon above the BF bound, since we are on a supersymmetric AdS$_3$ vacuum. This shows that the appearance of integer conformal dimensions is not necessarily a consequence of scale separation nor (extended) supersymmetry.

\subsection{No gauging and partial supersymmetry breaking}

Even though the superpotential \eqref{eq:W3D1} reproduces the correct scalar potential, one may wonder whether it provides the unique description, or whether there are further details that actually require a different one.
This is particularly important for us because the simultaneous presence of a superpotential and a D-term could lead to a vacuum that does not preserve the full amount of supersymmetry, but only part of it \cite{deWit:2003ja}.
This would therefore represent a loophole in the claim that our vacua preserve extended supersymmetry.

To show that this does not happen, we show that our superpotential \eqref{eq:W3D1} is indeed the only option.
Indeed, we know that the four-dimensional model is correctly described by a superpotential, so we just have to show that no D-term can give rise to the new flux contributions proportional to $f'$ and $h'$.

The complete scalar potential of three-dimensional $\mathcal{N}=2$ supergravity is \cite{deWit:2003ja}
\begin{equation}
V = V_{D} + V_{F} \,, 
\end{equation}
with the gauging piece given by
\begin{equation}
V_\text{D} = 4 \left( g^{I \bar J} \partial_I P \partial_{\bar J} P -  P^2 \right) \,, 
\end{equation}
where $P = \Theta^{AB} D_A D_B $. The $\Theta^{AB}$ are constants appearing in front of the Chern-Simons terms, while $D_A$ are the Killing moment maps whose derivative gives the killing vectors. 
The $V_F$ is the contribution from the superpotential.

Let us consider the DGKT reduced on S$^1$ with K\"ahler potential \eqref{eq:K3D1} and superpotential \eqref{eq:WDGKT}. 
Due to the isometries of the three-dimensional theory the only isometry we can gauge, if any, is 
\begin{equation}
\delta R = i \alpha \,,
\end{equation}
with constant killing vector and which leaves the K\"ahler potential invariant. 
The associated killing moment map is thus such that $\partial_{\bar R}D = -g_{R \bar R}$, giving 
\begin{equation}
D = \frac{1}{2(R + \bar R)} + \xi \,,
\end{equation}
where $\xi$ is a real constant, namely the Fayet-Iliopoulous term. 
The function $P$ takes then the form 
\begin{equation}
P = f' D^2 = f' \left( \frac{1}{2(R + \bar R)} + \xi \right)^2 \,,
\end{equation}
where the assumption $\Theta = f'$ is the only way to eventually obtain a coupling proportional to $(f')^2$. Indeed, the contribution that we get in the scalar potential from such $P$ is 
\begin{equation}
V_{D} = \frac{1}{(R + \bar R)^2} f'^2 \left( \frac{1}{2(R + \bar R)} + \xi \right)^2 \,. 
\end{equation}
This contribution to the potential depends only or $R$, whereas the $(f')^2$ term that appears in \eqref{eq:V3dDGKT} the dimensional reduction of $F_4 \wedge * F_4$ contains both $S$ and $R$. Hence, the only possible D-term fails in providing the correct contribution to the scalar potential. 
A similar argument works for the $(h')^2$ term coming from $H_3 \wedge * H_3$.

The only remaining option would be to gauge the R-symmetry.
Indeed, gauging $\mathrm{SO}(2)_R$ requires $W$ to vanish, since it transforms non-trivially under it \cite{deWit:2003ja}.
Conversely, the presence of a superpotential forbids the gauging of the R-symmetry.
In our model, a superpotential is always present, at least to describe the couplings inherited from the four-dimensional DGKT setup, and therefore R-symmetry gauging is excluded as well.
Finally, let us mention that our vacua are in fact $\mathcal{N}=(1,1)$ supersymmetric, since this is the only possibility when only a superpotential is present on the vacuum.

\section{Second model: CFI on a circle}

 The second model we present is a slight variant of the first. We consider massive type IIA supergravity compactified on an orientifold of the toroidal orbifold $T^6/\mathbb{Z}_2^2$, following CFI \cite{Camara:2005pr}. This construction gives rise to four-dimensional $\mathcal{N}=1$ AdS vacua, which provided a second example exhibiting scale separation. We further compactify on a circle so that the internal seven-dimensional space becomes 
\begin{equation}
\label{eq:Y7}
Y_7 = \frac{T^6}{\mathbb{Z}_2 \times \mathbb{Z}_2} \times S^1.
\end{equation}
The orientifold involution acts on the $T^6$ coordinates $z_i = x_i + i y_i$, $i=1,2,3$, as in \eqref{eq:sigmaO6} and leaves the $S^1$ direction invariant. 
The setup features the same fluxes as the previous one, and also the same parity \eqref{eq:sigmaH3F4} under the orientifold involution

As before, we first determine that the background admits three-dimensional AdS solutions with moduli stabilization and scale separation, and in a second step we show that such solutions preserve $\mathcal{N}=2$ supersymmetry. 
Due to the similarities with the previous model, we will be brief and mainly focus on the differences.

\subsection{Setup}

We start by giving the basis of harmonic three- and two-forms invariant under the orbifold involutions on which $\Omega$ and $J$ are expanded. The three-form basis is
\begin{equation}
\begin{aligned}
\alpha_0 &= dx_1 \wedge dx_2 \wedge dx_3\,, \qquad
& \beta_0 &= -dy_1 \wedge dy_2 \wedge dy_3\,, \\
\alpha_1 &=- dx_1 \wedge dy_2 \wedge dy_3\,, 
& \beta_1 &= dy_1 \wedge dx_2 \wedge dx_3\,, \\
\alpha_2 &=- dy_1 \wedge dx_2 \wedge dy_3\,, 
& \beta_2 &= dx_1 \wedge dy_2 \wedge dx_3\,, \\
\alpha_3 &=- dy_1 \wedge dy_2 \wedge dx_3\,, 
& \beta_3 &= dx_1 \wedge dx_2 \wedge dy_3\,,
\end{aligned}
\end{equation}
where $\alpha_0$, $\alpha_i$, $i=1,2,3$ are even under the orientifold involution, while $\beta_0$, $\beta_i$ are odd; notice that this is opposite to the DGKT example, but it amounts just to a different choice of complex structure. Indeed, recall that the orbifold $T^6/\mathbb{Z}_2^2$ has $h^{21}=3$, and in our conventions ${\rm Re} \Omega\propto \alpha_0, \alpha_i$ and ${\rm Im}\Omega \propto \beta_0, \beta_i$. The two-form basis is
\begin{equation}
\begin{aligned}
\omega_1 =& dx_1 \wedge dy_1\,, \\
\omega_2 =& dx_2 \wedge dy_2\,, \\
\omega_3 =& dx_3 \wedge dy_3,
\end{aligned}
\end{equation}
it is odd under the orientifold involution, and indeed the orbifold $T^6/\mathbb{Z}_2^2$ has $h^{11}=3$. 

Proceeding with the same logic as in the previous example, we then switch to the basis \eqref{eq:ytoxbasis} and introduce the three-form $\Phi$ 
\begin{equation}
\begin{aligned}
\Phi &= {\rm Im} \Omega + \eta \wedge J\\
&=-e^{456}- e^{135}  + e^{234} + e^{126} + e^{147} + e^{257} + e^{367}\\
&=\sum_{i=1}^7 s_i \Phi_i\, ,
\end{aligned}
\end{equation}
and the four-form $\Psi$
\begin{equation}
\begin{aligned}
\Psi &= \eta \wedge {\rm Re} \Omega -\frac12 J \wedge J \\
&= -e^{1237} -e^{2467}+e^{1567}+e^{3457}+e^{2356} + e^{1346}+ e^{1245} \\
&=*\Phi\, .
\end{aligned}
\end{equation}
A convenient basis of orientifold-odd and even forms is thus
\begin{align}
\Phi_i &=\{-dx^{456},-dx^{135},dx^{234},dx^{126},dx^{147},dx^{257},dx^{367}\},\\
\Psi_i &=\{-dx^{1237}, -dx^{2467}, dx^{1567}, dx^{3457},dx^{2356},dx^{1346},dx^{1245},\}\, .
\end{align}
To map to the previous example one sends $e^{3,6}_{DGKT} \to -e^{3,6}_{CFI}$.

We consider again the ten-dimensional string frame metric \eqref{eq:ds10d} and \eqref{eq:ds7d} and look for a minimal solution with only two independent moduli 
\begin{equation}
s_1=s_2=s_3=s_4 = r^3 \equiv s \, , \qquad s_5=s_6=s_7 = r^2 r_7 \equiv t \,, 
\end{equation}
such that the seven-dimensional volume in string units is ${\rm vol}(Y_7) = r^6 r_7 = s^{4/3} t$.

The untwisted cohomology of our seven-dimensional manifold \eqref{eq:Y7} has Betti numbers $b_1=1$, $b_2=3$ and $b_3 = 11$. The latter comes from the fact that with respect to DGKT we have $2 \times h^{21}=6$ additional three-cycles (recall $b_3 = 2 +2h^{21}$ for Calabi-Yau). Under the orientifold involution, these split as $b_3^- = 7$ and $b_3^+ = 4$, and thus $b_4^-=4$ and $b_4^+=7$ since the volume form is odd. For the same reasons as before, the manifold \eqref{eq:Y7} does not have $G_2$-holonomy nor $G_2$-structure. 
Furthermore, for the same reasons as in the previous example, we will expand $F_4$ on the harmonic even-form basis $\Psi_i$  and $H_3$ on the odd-form basis $\Phi_i$.

\subsection{Circle compactification and ten-dimensional solution}

To compactify CFI we need to switch-on additional RR-flux along the circle, avoiding that it shrinks leading us back to the four-dimensional AdS vacuum. We will follow very similar steps as in the DGKT circle compactification, so we will be rather brief and mainly explain the differences. 

We turn on a new contribution to the $F_4$ flux compatible with the $\mathbb{Z}_2^2$ orbifold and even under the orientifold involution. Due to the specific four-form basis we chose, the ansatz for $F_4$ is formally identical to \eqref{eq:F4dgkt+} and we repeat it below for convenience 
\begin{equation}
\label{eq:F4cfi+}
F_4 = f(\Psi_5+\Psi_6+\Psi_7)+f'( \Psi_1+\Psi_2+\Psi_3+\Psi_4)\, .
\end{equation}
The piece multiplied by $f$ is the usual CFI flux, while the piece multiplied by $f'$ is the new contribution proportional to $\alpha_0\wedge dx^7$, $\alpha_i\wedge dx^7$. A difference with respect to the DGKT example is that now all seven monomials in \eqref{eq:F4cfi+} can in principle have independent coefficients, since $Y_7$ has $b_4^{+}=7$ different orientifold-even four-cycles. Still, for simplicity we consider the isotropic solution in which only $f$ and $f'$ are different.  

To avoid introducing $O2/D2$-sources and yet solve the $dF_6$ tadpole, we ask once more that
\begin{equation}
    F_4 \wedge H_3 =0\, ,
\end{equation}
and thus we employ an ansatz for $H_3$ formally identical to \eqref{eq:H3dgkt+}, that we repeat below for convenience 
\begin{equation}
H_3 = h (\Phi_1+\Phi_2+\Phi_3+\Phi_4) + h'(\Phi_5+\Phi_6+\Phi_7)\, .
\end{equation}
Here, $h$ is the usual CFI flux while $h'$ is a new contribution proportional to $\omega_i \wedge dx^7$. 
As for $F_4$, all seven monomials in the expression above can in principle scale independently since the manifold $Y_7$ has $b_3^-=7$; this is once more a difference with respect to the DGKT setup. However, for simplicity we choose to keep only $h$ and $h'$ as independent flux parameter in the solution. 

The discussion on the tadpole cancellation conditions proceeds now exactly as in the previous example. After the newly introduced fluxes, the only non-trivial relation is the $dF_2$ tadpole. To cancel it and to avoid breaking supersymmetry further, we introduce $D6$-branes in a mutually supersymmetric way with the orientifold background, as displayed in the table \ref{table:D6O6conf}. The $dF_2$ tadpole becomes thus \eqref{eq:dF2tad} and we also have the constraint \eqref{eq:fh}. The same discussion on the signs \eqref{eq:hh'sign}, \eqref{eq:ff'sign} and \eqref{eq:hmsign} applies here. 
We will also assume $|f| \sim |f'|$ but, as explained, this setup allows now for more freedom of non-isotropic scalings and fluxes and it is conceivable that larger classes of solutions exists, even if we do not explore them here.

\subsubsection*{Ten-dimensional solution}
At this point, thanks to our basis and flux choices, we have reached exactly the same configuration as in the previous example so that all of our previous steps in section \ref{sec:10dsolDGKT} can be repeated without any modification (except for the twisted sector) and a solution of ten-dimensional bispinor equations \eqref{eq:eomH3}-\eqref{eq:bispinorpairing}, and thus of \eqref{eq:eom1}-\eqref{eq:eom6}, is found. This solution has exactly the same form as in \eqref{eq:sol10ds}-\eqref{eq:sol10dgs} and, crucially, it features parametric large volume, weak coupling and scale separation as in \eqref{eq:scalesep}. It also comes in two branches depending on the sign of $m$. To establish that this solution preserves not only $\mathcal{N}=1$ supersymmetry but actually $\mathcal{N}=2$, we look at the three-dimensional effective theory. However, since the equations of motion involved are exactly the same as in the previous setup, where we have already shown that they preserve  $\mathcal{N}=2$ supersymmetry, we really do not need to check it once more. Still, we present the three-dimensional theory for completeness.

\subsection{The three-dimensional effective theory}

The four-dimensional $\mathcal{N}=1$ supergravity describing massive type IIA string theory on the orientifold of the toroidal orbifold $Y_6 = T^6/\mathbb{Z}_2^2$ contains seven chiral multiplets, $S$, $T_i$ and $U_i$ with $i={1,2,3}$ \cite{Derendinger:2004jn,Camara:2005pr}. There are also forty-eight additional chiral multiplets parametrizing the twisted sector, see $e.g.$ \cite{Denef:2005mm}, but we not consider them in this section. The chiral multiplets $S$ and $U_i$ arise form the complex structure sector of the parent four-dimensional $\mathcal{N}=2$ theory with $h^{21}=3$ and contain the (four-dimensional) dilaton, while $T_i$ describe the geometric moduli of the K\"ahler sector with $h^{11}=3$. 

 We denote ${\rm Re}T_i = r_ir_{i+3}$, $i=1,2,3$, the volume of the two-cycles in string units, ${\rm Re} S = e^{-\phi} \sqrt{{\rm Vol}(Y_6)}/\sqrt{\tau_1 \tau_2 \tau_3}$, with ${\rm Vol}(Y_6) = \prod_{i=1}^3 r_i r_{i+3}$ and $\tau_i = r_{i+3}/r_i$, and furthermore ${\rm Re} U_i = e^{-\phi}\sqrt{{\rm Vol}(Y_6)}\sqrt{\frac{\tau_j \tau_k}{\tau_i}}$, $i \neq j \neq k$. We also set $t_i = r_i r_{i+3}r_7 =s_{i+4}$, $i=1,2,3$, such that  ${\rm Re}T_i = t_i/r_7$. The imaginary parts needed to complete these multiplets are provided by axions arising from the reduction of $B_2$ on the $b_2^{CFI,-}=3$ two-cycles and of $C_3$ on the $b_3^{CFI,+}=4$ three-cycles. Eventually, the K\"ahler potential and the superpotential of the four-dimensional supergravity read (up to a real constant) \cite{Derendinger:2004jn,Camara:2005pr}
 \begin{align}
 K_{CFI} & = - \sum_{i=1}^3 \log (T_i + \bar T_i) -  \log (S+\bar S) - \sum_{i=1}^3 \log (U_i + \bar U_i)\, ,\\
 W_{CFI} &=  \frac i2 h (S+ \sum_{i=1}^3U_i) - \frac i2 f \sum_{i=1}^3 T_i + 2 i m T_1 T_2 T_3 \, .
 \end{align}

The string frame metric we consider for the whole reduction is
\begin{equation}
\begin{aligned}
ds_{10}^2 &= \frac{e^{2\phi}}{{\rm Vol}(Y_6)}ds_4^2 + ds_6^2\\
&=\frac{e^{2\phi}}{{\rm Vol}(Y_6)}\left(\frac{1}{{\rm Re}\, R} ds_3^2 +{\rm Re}\, R\,\, dx_7^2\right)+ ds_6^2\, ,
\end{aligned}
\end{equation}
where $R$ is a new chiral superfield with ${\rm Re}\, R = (r_7 {\rm Re}S)^2 \tau_1 \tau_2 \tau_3$. 
The three-dimensional K\"ahler potential is then
\begin{equation}
K = -\log (R+ \bar R)- \sum_{i=1}^3 \log (T_i + \bar T_i) -  \log (S+\bar S) - \sum_{i=1}^3 \log (U_i + \bar U_i)\, ,
\end{equation}
however to find the superpotential $W$ is non-trivial. One can check that simply replacing $S \to S + \sum_{i=3}U_i$ into \eqref{eq:W3D1} in order to preserve the three isometries of CFI, associated to the three massless axions in four dimensions, does not reproduce the scalar potential obtained from dimensional reduction. The correct superpotential was provided to us by \texttt{ChatGPT} using \texttt{GPT-5.5 Thinking} and it reads\footnote{The prompt that \texttt{ChatGPT} received from us consisted in a previous draft of the paper and a plausible ansatz for $W$, which however we knew to be not completely correct. \texttt{ChatGPT} properly understood the issue and came up with the correct formula after roughly 6 minutes of \emph{thinking}. When asked whether or not the result was obtained by using a particular reference, \texttt{ChatGPT} answered: ``I did not recognize it from a standard reference. I inferred it by reverse-engineering a holomorphic ansatz whose F-term potential reproduces your direct-reduction target.'' It thus seems that \texttt{ChatGPT} was able to obtain a genuinely new and non-trivial formula after our input.}
\begin{equation}
\begin{aligned}
\label{eq:W3D2GPT}
W &=  \frac i2 h \left(S+\sum_{i=1}^3 U_i\right)
-\frac{i}{2} f \sum_{i=1}^3 T_i
+2 i m T_1 T_2 T_3\\[1mm]
&\quad
+i\frac{h'}{2}\sqrt{S U_1 U_2 U_3}
\left[
\sqrt{\frac{T_2 T_3}{R T_1}}
+\sqrt{\frac{T_1 T_2}{R T_3}}
+\sqrt{\frac{T_3 T_1}{R T_2}}
\right]\\[1mm]
&\quad
-\frac{i f'}{2}\sqrt{S U_1 U_2 U_3}
\left(
\frac{1}{S}+\sum_{i=1}^3\frac{1}{U_i}
\right)
\sqrt{\frac{T_1 T_2 T_3}{R}} \, \,.
\end{aligned}
\end{equation}
The second and third line is the new contribution to CFI due to the circle compactification. It features a function of the fields $S$, $U_i$ which is highly non-trivial to guess as a generalization of the previous example \eqref{eq:W3D1}. This superpotential explicitly breaks the original CFI isometries due to terms proportional to $1/R$. Indeed, the combination $S+ \sum_{i=1}^3 U_i$ allows for three independent shifts of the imaginary parts of $U_i$ each of which can be compensated by a shift of the imaginary part of $S$. This property is explicitly spoiled by the couplings in the second and third line. Thus, we expect the three-dimensional model to achieve full (untwisted) moduli stabilization at the classical level, contrary to the four-dimensional CFI setup we started from. We will verify this expectation shortly by explicitly computing the spectrum on the vacuum. 

The ten-dimensional scalar potential coming from dimensional reduction and with vanishing axions is
\begin{equation}
\begin{aligned}
\label{eq:V3dCFI}
\left(256 \,\frac{ (t_1 t_2 t_3)^{2/3}}{(s_1 s_2 s_3 s_4)^{-2/3}g_s^{6}}\right) V &= 16 m^2\\
&+\frac{ h^2}{g_s^2}\left(\frac{1}{s_1^2}+\frac{1}{s_2^2}+\frac{1}{s_3^2}+\frac{1}{s_4^2}\right) + \frac{ (h')^2}{g_s^2}\left(\frac{1}{t_1^2}+ \frac{1}{t_2^2}+\frac{1}{t_3^2}\right)\\
&+\frac{f^2}{(s_1 s_2 s_3 s_4)^{2/3} (t_1 t_2 t_3)^{2/3}}\left(t_1^2 + t_2^2 + t_3^2\right) \\
&+ \frac{ (f')^2 }{(s_1 s_2 s_3 s_4)^{2/3}(t_1 t_2 t_3)^{2/3}}(s_1^2+s_2^2+s_3^2+s_4^2)\\
&- \frac{8hm}{g_s}\left(\frac{1}{s_1}+\frac{1}{s_2}+\frac{1}{s_3}+\frac{1}{s_4}\right)-\frac{8 h' m}{g_s}\left(\frac{1}{t_1}+\frac{1}{t_2}+\frac{1}{t_3}\right)\\
&+\frac{2}{g_s (s_1 s_2 s_3 s_4)^{1/3}(t_1 t_2 t_3)^{1/3}}\left(3 fh' + 4f'h\right).
\end{aligned}
\end{equation}
We compare it with the couplings arising from
\begin{align}
F_0 \wedge * F_0 &= F_0^2 {\rm vol}_7\, ,\\
e^{-2\phi} H_3 \wedge * H_3 &= e^{-2\phi}{\rm vol}_7 \left[h^2\left(\frac{1}{s_1^2}+\frac{1}{s_2^2}+\frac{1}{s_3^2}+\frac{1}{s_4^2}\right) +(h')^2\left(\frac{1}{t_1^2}+\frac{1}{t_2^2}+\frac{1}{t_3^2}\right)\right]\, ,\\
F_4 \wedge * F_4 &= \frac{{\rm vol}_7}{(s_1 s_2 s_3 s_4)^{2/3} (t_1 t_2 t_3)^{2/3}}\left[f^2 (t_1^2+t_2^2+t_3^2) + (f')^2 (s_1^2+s_2^2+s_3^2+s_4^2) \right]\, ,
\end{align}
and the source actions in the smeared approximation
\begin{align}
S_{DBI}^{O6,i} &\simeq \int \frac{e^{-\phi}}{s_i}{\rm vol}_7\, , \qquad i=1,2,3,4\,,\\\
S_{DBI}^{D6,i} &\simeq \int \frac{e^{-\phi}}{t_i}{\rm vol}_7\, , \qquad i =1,2,3\, .
\end{align}
One can see that the first line in \eqref{eq:V3dCFI} comes from $F_0 \wedge * F_0$, the second line from $H_3 \wedge * H_3$, the third and fourth line from $F_4 \wedge * F_4$, the fifth line from the smeared sources and the last line vanishes due to the tadpole condition \eqref{eq:fh}. We believe that this matching provides non-trivial evidence that the superpotential \eqref{eq:W3D2GPT} provided by \texttt{ChatGPT} gives the correct description of the ten-dimensional model.

We can now look for AdS vacua of the three-dimensional theory by solving the F-term conditions. 
A solution is given by ($T_i \equiv T$, $U_i \equiv S$ for $i=1,2,3$, $i.e.$ $ s_i \equiv s$ and $t_i \equiv t$)
\begin{align}
S = U = \frac{3 \sqrt 7}{64} f \sqrt\frac fm\, , \qquad T = \frac{\sqrt 7}{4} \sqrt \frac fm\, , \qquad R = \frac{567 \sqrt 7}{256}\frac{f^2 (h')^2}{h^4}\left(\frac fm\right)^\frac32 \, ,
\end{align}
or in ten-dimensional language
\begin{equation}
s = \frac{7^\frac34}{8}\left(\frac fm\right)^\frac34 \,,\qquad t= 4\cdot 7^\frac34 \frac{f'}{m}\left(\frac mf\right)^\frac14 \, , \qquad g_s = \frac83 7^\frac14 \frac hm \left(\frac mf\right)^\frac34\, .
\end{equation}
The supersymmetric vacuum energy is 
\begin{equation}
V = - 4 e^K W \overline W = -\frac{2^{12}}{3^8 7^2} \frac{h^8 m^4}{f^8 (h')^2}\, ,
\end{equation}
with $W = \frac{7 \sqrt 7 }{64}i f \left(\frac fm\right)^\frac12$. We have thus found an AdS vacuum with $\mathcal{N}=2$ supersymmetry. It arises from the same scalar potential of the ten-dimensional model and the solution matches with \eqref{eq:sol10ds}-\eqref{eq:sol10dgs} up to a trivial rescaling of the fields, $s \to s/(2 \sqrt 2)$, $t \to \sqrt 2 t$, $g_s \to 2 \sqrt 2 \, g_s$. 
We also evaluated the full Hessian matrix on this specific solution and we found that it has non-vanishing determinant, implying full moduli stabilization. Recall, however, that we are once more restricting to the untwisted and closed string sector only. Finally, the conformal dimensions of the operators of the putative dual two-dimensional field theory are not integer, namely
\begin{equation}
\begin{aligned}
1 + \sqrt{1+ m^2 L_{AdS}^2} \simeq &\left\{1.43, 1.43, 1.43, 2.43, 2.43, 2.43, 3.57, 3.57,\right.\\
&\, \, \, \left. 4.57, 4.57, 1.90, 3.18, 8.94, 2.9, 4.18, 7.94\right\}.
\end{aligned}
\end{equation}

\section{Outlook}

In this work, we provided the first examples of scale-separated vacua with extended supersymmetry.
They arise from compactifications of massive type IIA supergravity on a seven-dimensional toroidal orbifold with fluxes, O-planes, and D-branes.
The vacua are purely classical and achieve parametric scale separation with full moduli stabilization at weak string coupling and large internal volume.

Let us comment on the assumptions and limitations of our ten-dimensional construction.
We assumed smeared sources for both O-planes and D-branes.
Until recently, this was motivated by the absence of suitable techniques to deal with the equations of motion with localized sources, except in specific situations.
In \cite{Junghans:2020acz,Marchesano:2020qvg}, a general and systematic procedure was put forward to capture corrections to the smeared approximation directly from the ten-dimensional point of view.
It would be of clear interest to apply this technique to our setup and go beyond the assumption employed here.

Relatedly, a particular concern is the possible presence of intersecting sources even after the orbifold singularities are blown up; we refer to \cite{Junghans:2023yue} for a detailed discussion of this issue with explicit examples.
In the same work, it is shown that the intersection of the O-planes in DGKT is removed when passing from the orbifold to a smooth Calabi--Yau.
To place our construction on firmer footing, it would be important to perform a similar analysis for CFI.
Moreover, the twisted sector clearly deserves further investigation, in both models and in both three and four spacetime dimensions.

More generally, we performed only a preliminary analysis of the setups under investigation, which was nevertheless sufficient to find two explicit solutions.
It seems plausible that further solutions exist when considering less isotropic ans\"atze for the fluxes and radii.
Another valuable application would be to study these vacua via holography, since two-dimensional conformal field theories are typically better understood than three-dimensional ones.

Finally, we reiterate that parts of one of the three-dimensional superpotentials appearing in this work was obtained by \texttt{ChatGPT}, namely formula \eqref{eq:W3D2GPT}.
Its dependence on the scalar fields seems to us highly non-trivial to guess solely by comparison with known expressions.
This specific formula does not by itself represent the main result of our work, nor would its absence have jeopardized our analysis. 
Nevertheless, this fact shows that Large Language Models have now reached a level at which, upon proper input, they can genuinely find new results.

\appendix

\section*{Acknowledgements}

We would like to thank G. Tringas, V. Van Hemelryck and T. Van Riet for discussions. The work of NC is supported by the Research Foundation Flanders (FWO grant 1259125N). NC thanks the University of Hamburg for kind hospitality while part of this work was completed.

\bibliography{references}  
\bibliographystyle{utphys}

\end{document}